\documentclass[final,5p,times,twocolumn]{elsarticle}

\usepackage{amsmath}
\usepackage{amssymb}
\usepackage{booktabs}
\usepackage{multirow}
\usepackage{array}
\usepackage{tabularx}
\usepackage{graphicx}
\usepackage{siunitx}
\usepackage{xcolor}
\usepackage{microtype}
\usepackage{url}
\usepackage{placeins}

\journal{Energy}

\newcommand{\CA}{^\circ\mathrm{CA}}
\newcommand{\aTDC}{^\circ\mathrm{aTDC}}

\begin{document}
\raggedbottom

\begin{frontmatter}

\title{Methanol substitution suppresses deflagration-excited knock and increases the efficiency of a hydrogen-fuelled CFR engine}

\author[aff1]{Riming Xu}
\author[aff1]{Mebin Panithasan}
\author[aff1]{Xinyang Wang\corref{cor1}}
\author[aff1]{Giovanni Vorraro}
\author[aff1]{James W.G. Turner}
\cortext[cor1]{Corresponding author. Email: xinyang.wang@kaust.edu.sa}
\address[aff1]{Clean Energy Research Platform, King Abdullah University of Science and Technology (KAUST), Thuwal 23955-6900, Saudi Arabia}

\begin{abstract}
Knock restricts the performance of hydrogen spark-ignition engines and is usually attributed to end-gas autoignition. To the authors' knowledge, this is the first study in which methanol is used to replace part of the hydrogen in a hydrogen engine at constant fuel energy addition, and it shows that near the knock boundary the limiting oscillation is excited by the deflagration itself, which methanol removes without derating the performance. Port-injected methanol replaced 20--50\% of the hydrogen energy in a variable-compression Cooperative Fuel Research engine at 600 rpm and compression ratios of 8--12. With pure hydrogen, oscillation in the 4--15 kHz range began before 10\% of the charge had burned in 97--100\% of cycles, and its maximum amplitude (MAPO) followed the maximum pressure-rise rate across 2499 cycles (Spearman $\rho=0.94$). Methanol slowed the deflagration: it replaced high-flame-speed hydrogen, cooled the intake by 24--25 K and, at unchanged intake pressure, enleaned the mixture by 10--13\%. At 50\% methanol, the 10--90\% combustion duration lengthened 2.7--3.0-fold, oscillation onset moved 4.5--8.3$\CA$ after CA10 and mean MAPO fell by 89--93\% although the spark timing was advanced by 13--19$\CA$, and IMEP and indicated thermal efficiency rose by 2.4--7.8\%. Methanol was thus worth more than this advance, and at matched CA50 MAPO was still about 90\% lower. The measured combustion duration of all 25 data sets scaled with the computed laminar flame speed as $S_L^{-0.58}$ ($R^2=0.99$); fuel replacement accounted for about three quarters of the reduction in $S_L$, and the end-gas autoignition integral stayed below 0.07, against 1 at autoignition. The chamber kept its cylindrical acoustic modes, whose frequencies fell by 8--9\%. The trade-off is formaldehyde emissions, which rose five- to sixfold while NO$_x$ fell by 49--85\%.
\end{abstract}

\begin{highlights}
\item Near the H$_2$ knock limit, oscillation began before CA10 in 97--100\% of cycles.
\item MeOH enabled over 13--19$^\circ$ more spark advance; knock fell 90\% at equal CA50.
\item Port-injected MeOH cooled the intake by 24 K and enleaned the charge passively.
\item At 50\% MeOH, MAPO fell by 89--93\% while indicated efficiency rose by 2--8\%.
\item Chamber acoustic modes were unchanged; their frequencies fell 8--9\% with MeOH.
\end{highlights}


\end{frontmatter}

\section{Introduction}
\label{sec:introduction}

Hydrogen-fuelled internal combustion engines offer a carbon-free route for heavy-duty, off-road and marine applications that are difficult to electrify, while reusing established engine hardware and manufacturing capacity \cite{Verhelst2009hydrogen,Zhou2025abnormal}. Hydrogen's wide flammability range and high laminar burning velocity permit lean, unthrottled operation, and dedicated hydrogen engines have reached peak brake thermal efficiencies of about 45\% at automotive-relevant cylinder capacities \cite{Verhelst2009hydrogen,Vancoillie2012comparison}. The same properties, together with a low minimum ignition energy, make hydrogen prone to abnormal combustion, including backfire in port-injected engines, hot-spot-induced pre-ignition and knock \cite{Gao2022review,Golc2025hot,Zhou2025abnormal}. Knock restricts the usable equivalence ratio, compression ratio, and spark advance and therefore caps the torque and efficiency of hydrogen engines \cite{Szwaja2007comparisons,Wang2024working}. Hydrogen is assigned a methane number of zero \cite{Verhelst2009hydrogen}, and the standard research octane number (RON) procedure rates it at only 62--64, although that procedure requires very rich operation that is not representative of practical hydrogen engines \cite{Poursadegh2023autoignition}. Other studies have reported an effective RON above 130 and a motor octane number (MON) of 60 \cite{Verhelst2009hydrogen}, so the knock resistance attributed to hydrogen depends strongly on the rating method and mixture strength.

In hydrogen engines the term `knock' covers more than one phenomenon. It conventionally denotes end-gas autoignition, but what is measured is an abnormal oscillation of cylinder pressure, which can also result from other events such as surface ignition \cite{Verhelst2009hydrogen,Wang2017knocking}. Szwaja and Naber distinguished light hydrogen knock, generated by fast and unstable combustion from the start of combustion with pressure-pulsation amplitudes of about 20--100~kPa, from heavy knock caused by autoignition of the unburnt mixture with amplitudes above 200~kPa \cite{Szwaja2013dual}. Deflagration-based knock without end-gas autoignition has likewise been reported for stoichiometric methanol combustion, with a lower maximum amplitude of pressure oscillation (MAPO) than autoignition-based knock \cite{Singh2024deflagration}. More generally, the excitation of chamber resonance scales with the local pressure gradients created by heat release, and smooth flame propagation excites it only weakly \cite{Guardiola2018cylinder,Vressner2003pressure}. In a hydrogen direct-injection Cooperative Fuel Research (CFR) engine, high flame speed, combustion instability and end-gas autoignition were all found to generate pressure oscillations, making amplitude-only knock diagnosis unreliable \cite{Wang2026characterization}. Pressure-based metrics such as MAPO, the maximum pressure-rise rate (MPRR) and integrated knock intensities \cite{Brunt1998gasoline,Naber2006combustion,Kalghatgi2017knock} are not interchangeable: in CFR engines, pressure-rise-rate metrics did not correlate directly with the standard knockmeter reading \cite{Rockstroh2018insights}, and different metrics selected different maximum-knock cycles \cite{Wang2025comparative}. The oscillation is carried by the chamber acoustic modes whose frequencies scale with the speed of sound and bore diameter \cite{Draper1938pressure,Scholl1998volume}, and hydrogen knock modes lie at slightly higher frequencies than those of gasoline \cite{Szwaja2007comparisons,Luo2016inducing}. Because knock intensity behaves as a cyclically independent random variable \cite{Spelina2013characterization}, a knock-mitigation strategy is better judged from amplitude, rate, spectral and cycle-distribution measures together than from a single condition-mean value \cite{Dieguez2018characterization}.

Knock in hydrogen engines is usually avoided by leaner operation or retarded spark timing, both of which cost power density or efficiency \cite{Hong2024progressive,Purayil2024spark,Purayil2024influence}. Exhaust gas recirculation and charge dilution reduce knock propensity, with recirculation more effective than enleanment for hydrogen \cite{Szwaja2019dilution}, and in direct-injection engines mixture stratification, injection timing and valve timing provide further control \cite{Hong2024progressive,Liang2025suppressing}. Water injection extended the knock limit of a boosted heavy-duty hydrogen engine enough to restore optimal spark timing at richer mixtures; kinetic modelling attributed the benefit mainly to charge cooling, with a smaller thermophysical effect and a weak kinetic effect \cite{Mortimer2023extending}. Port water injection has similarly removed knock in hydrogen-fuelled argon power cycle engines \cite{Jin2023knock}, and, conversely, increasing the intake temperature reduces the hydrogen knock limit \cite{Purayil2025impact}. Blending ammonia slows the hydrogen flame and suppresses knock \cite{Ge2023knock,Hong2024ammonia}. These measures either add an energy-free fluid that must be stored and metered separately, or trade knock margin for slower combustion or reduced load. A co-fuel that cools the charge and resists autoignition while also supplying energy could avoid this derating penalty.

Methanol is a natural candidate for such a co-fuel. It can be synthesized from renewable hydrogen and captured CO$_2$, and hydrogen-derived methanol compares favourably with compressed or liquid hydrogen in production cost and well-to-tank efficiency \cite{Vancoillie2012comparison,Verhelst2019methanol}. Its heat of vaporization, 1100~kJ~kg$^{-1}$, is three to six times that of gasoline; it also has a high octane rating, a high flame speed and a low stoichiometric air--fuel ratio \cite{Verhelst2019methanol,Wang2019methanol}. Alcohols resist knock through both charge cooling and autoignition chemistry \cite{Westbrook2017chemical,Wang2017ethanol}, in-cylinder thermometry has quantified the evaporative cooling of alcohol blends \cite{Williams2014high}, and dual-injection strategies have used methanol and ethanol to suppress knock on demand \cite{Daniel2012dual,Huang2021dual}. When methanol is injected into the intake, most of the cooling comes from evaporation of the liquid film on the intake walls rather than from droplets in flight, as measured and modelled on a CFR engine \cite{Pu2024modelling}. Methanol engines nevertheless knock at high compression ratio \cite{Duan2022knock}, and stoichiometric methanol combustion can show deflagration-based knock \cite{Singh2024deflagration}.

Most hydrogen--methanol engine studies have used hydrogen as a minor additive in a methanol engine. Small hydrogen fractions extend the lean-burn limit, stabilize combustion and allow later ignition \cite{Ji2013enhancing,Zhang2014combustion,Gong2019influence,Gong2020evaluation,Zhen2020comparative}, and hydrogen pre-chamber ignition has enabled ultra-lean methanol combustion in a heavy-duty engine \cite{Uddeen2026pre}. Methanol and hydrogen have also been compared as separate fuels in the same spark-ignition engine \cite{Catapano2024complete}. The reverse configuration, in which methanol relieves the knock limit of a hydrogen-dominant engine, has received little attention. The closest studies blended methanol into gasoline and showed that the blend increased the hydrogen flow rate tolerated before knock, by 42.8\% for an M50 blend injected at 140~bar, by cooling the end gas \cite{Purayil2024influence,Purayil2026optimizing}; water--methanol injection has also been combined with turbocharging in a hydrogen engine, mainly to reduce emissions \cite{Chitragar2023influence}. In the methanol--gasoline studies, hydrogen was a secondary fuel added in steps until knock was detected, total fuel energy was not held constant, and knock was reported as the hydrogen flow rate at which it first appeared, with no measure of its intensity or of its timing within the cycle. Methanol also introduces pollutant species absent from hydrogen combustion: methanol engines emit formaldehyde and unburned methanol, particularly at cold start \cite{Wei2009formaldehyde,Li2009emissions,Zhou2024application}, hydrogen addition alters these unregulated emissions \cite{Gong2024computational}, and formaldehyde measured by Fourier-transform infrared (FTIR) spectroscopy can exceed chromatographic results because of spectral interference unless methanol-specific calibration is applied \cite{Geng2015comparative,Kraus2022adaptation,Gainey2025towards}.

Variable-compression CFR engines provide a controlled platform for these questions. They have been used to characterize and rate hydrogen knock \cite{Szwaja2007comparisons,Poursadegh2023autoignition}, to analyse hydrogen combustion versus compression ratio and mixture strength \cite{Beccari2023combustion}, and to compare knock metrics and the knock severity of alcohol blends \cite{Rockstroh2018insights,CorralGomez2020comparative,Wang2025comparative}. To the authors' knowledge, the present study is the first in which methanol replaces part of the hydrogen in a hydrogen engine at constant fuel power. It tests two linked hypotheses on a variable-compression CFR engine. First, that near the hydrogen knock boundary the limiting pressure oscillation is excited by the deflagration itself rather than by end-gas autoignition. Second, that port-injected methanol suppresses this oscillation at constant fuel power by slowing heat release via evaporative charge cooling, the leaner mixture that cooling and reduced hydrogen displacement produce, and the replacement of high-flame-speed hydrogen by a lower-flame-speed fuel, all of which help to avoid derating. Starting from near-boundary pure-hydrogen states at compression ratios of 8--12, methanol replaced 20--50\% of the hydrogen energy at constant commanded fuel power, with the air flow free to vary and the spark timing re-optimized. The analysis locates oscillation onset, the MPRR and the MAPO within each cycle's combustion, quantifies the cooling and enleanment from intake measurements, bounds the spark margin that methanol provides, compares knock at matched combustion phasing, relates the measured burn rate and knock to computed laminar flame speeds and end-gas autoignition, and identifies the chamber acoustic modes, so that the suppression is explained rather than only measured.

\section{Experimental and analytical methods}
\label{sec:methods}

\subsection{Engine, fuel delivery and data acquisition}
\label{sec:engine}

\begin{figure*}[tp]
\centering
\includegraphics[width=\textwidth]{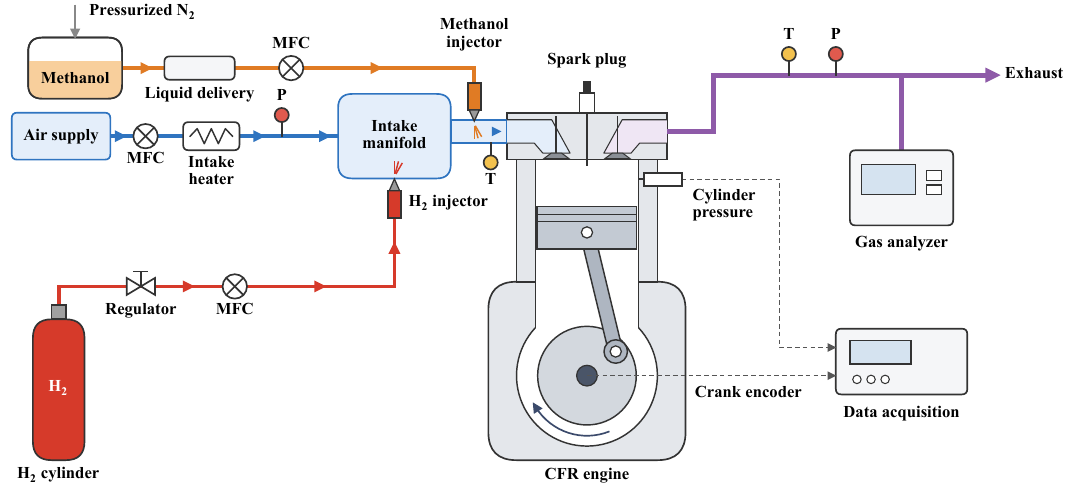}
\caption{CFR engine and measurement system. Hydrogen and methanol are supplied separately to the intake manifold through port injectors; the intake temperature is measured downstream of the methanol injector, and cylinder pressure is recorded against crank angle. MFC: mass-flow controller; P: pressure; T: temperature.}
\label{fig:setup}
\end{figure*}

Experiments were conducted in a modified single-cylinder Waukesha CFR F1/F2 variable-compression engine at 600 rpm (Fig.~\ref{fig:setup}; Table~\ref{tab:setup}). Bore, stroke and connecting-rod length are 82.55, 114.3 and 256 mm, giving a swept volume of 611.7 cm$^3$. The engine was operated at compression ratios (CR) of 8--12.

Dry intake air was metered by a Brooks SLA5800 thermal mass-flow meter. During methanol substitution its setting was not changed, so the air flow was set by the engine's own breathing; the intake pressure at inlet-valve closure stayed between 0.992 and 0.998 bar in all 25 data sets. Hydrogen was regulated to approximately 6 bar, metered by a second Brooks SLA5800 and delivered from $-280\aTDC$ through a Bosch EV14 NGI2 port-fuel injector. Methanol was supplied from a nitrogen-pressurized (approximately 4 bar) circuit through a Bronkhorst liquid mass-flow controller to a VW5405 injector in the intake manifold, starting at approximately $-300\aTDC$. The intake-temperature thermocouple is located in the intake manifold downstream of the methanol injector, whereas the air-heater and intake-tube thermocouples are upstream of it; a further thermocouple monitors the injector holder. Cylinder pressure was measured by an AVL QC34D transducer installed through the detonation-pickup D-1 cavity using an adapter and was sampled every $0.1\CA$ against an AVL 365C encoder, i.e.\ at 36 kHz at 600 rpm. Each data set contains 100 consecutive cycles.

\begin{table*}[tp]
\caption{Engine, fuel-delivery and data acquisition parameters.}
\label{tab:setup}
\centering
\begin{tabularx}{\textwidth}{@{}llX@{}}
\toprule
Category & Quantity & Value or definition \\
\midrule
Engine & Configuration & Modified single-cylinder Waukesha CFR F1/F2; two valves \\
 & Bore $\times$ stroke; connecting rod & 82.55 mm $\times$ 114.3 mm; 256 mm \\
 & Swept volume & 611.7 cm$^3$ \\
 & Compression ratio & 8, 9, 10, 11 and 12 \\
 & IVO / IVC; EVO / EVC & $-350$ / $-146\aTDC$; $140$ / $-345\aTDC$ \\
 & Speed & 600 rpm \\
\addlinespace
Acquisition & Crank-angle resolution & $0.1\CA$ (36 kHz at 600 rpm); 100 cycles per data set \\
 & Cylinder-pressure transducer & AVL QC34D through the D-1 cavity \\
 & Crank-angle encoder & AVL 365C \\
 & Intake and exhaust pressure & AVL GH14D absolute pressure sensors in the intake and exhaust manifolds \\
 & Intake temperature & Thermocouple in the intake manifold downstream of the methanol injector \\
\addlinespace
Fuel system & Air & Brooks SLA5800 thermal mass-flow meter; setting unchanged during substitution \\
 & Hydrogen & Approximately 6 bar; Brooks SLA5800; Bosch EV14 NGI2 port-fuel injector \\
 & Methanol & N$_2$-pressurized to approximately 4 bar; Bronkhorst liquid mass-flow controller; VW5405 port-fuel injector \\
 & Exhaust analyser & FTIR exhaust gas analyser reporting HCHO, NO$_x$, CO and hydrocarbons as wet-basis volume fractions (ppm), without a CH$_3$OH channel \\
\bottomrule
\end{tabularx}
\end{table*}

\subsection{Test campaigns}
\label{sec:campaigns}

A dry-hydrogen mapping campaign first defined the knock context. Throughout, `dry' and `dry-H$_2$' refer only to this campaign, in which no methanol was injected. At each CR the mixture was enriched step by step, with the spark advance retimed at every step so that CA50 stayed as close as practicable to $6\aTDC$; near the richest achievable mixture, the spark timing was then swept at constant $\lambda$. In total, 107 sets of 100 cycles each were processed.

The substitution campaign reproduced one near-boundary pure-hydrogen state at each CR and then replaced 20, 30, 40 and 50\% of the hydrogen energy with methanol (Table~\ref{tab:matrix}). Hydrogen and methanol target flows were chosen to keep the commanded fuel power equal to that of the pure-hydrogen state, the air setting was left unchanged, and spark timing was re-optimized at every step. The campaign comprised 25 data sets and 2500 cycles. One CR12 pure-hydrogen cycle ignited before the spark (pre-ignition; Supplementary Fig.~S6, red trace); because its spark-referenced MAPO, CA10, CA50 and CA90 do not describe knock, it was excluded from all statistics, which therefore rest on 2499 cycles. Because the mixture strength followed from the substitution itself and spark timing was re-optimized, each step measures the combined effect of the substitution at equal commanded fuel power; the separate contributions of phasing and of fuel replacement, cooling and enleanment are deconflated in Sections~\ref{sec:marginmethod} and \ref{sec:kinmethod}.

\begin{table*}[tp]
\caption{Test conditions of the substitution campaign. Each CR series comprises five data sets at 0, 20, 30, 40 and 50\% methanol energy fraction. Mixture-wide $\lambda$ is calculated from Eq.~\eqref{eq:mixedlambda} with measured air and hydrogen flows; intake temperature is measured downstream of the methanol injector. Spark timings are listed for 0/20/30/40/50\% methanol.}
\label{tab:matrix}
\centering
\small
\setlength{\tabcolsep}{5pt}
\begin{tabular}{ccccc}
\toprule
CR & \shortstack{Spark timing\\($\aTDC$)} & \shortstack{$\lambda_{\mathrm{mix}}$\\0\% $\rightarrow$ 50\%} & \shortstack{Intake temperature\\0\% $\rightarrow$ 50\% ($^\circ$C)} & Cycles \\
\midrule
8 & 0.5/$-4.5$/$-8.5$/$-9.5$/$-12.5$ & 1.20 $\rightarrow$ 1.34 & 27.3 $\rightarrow$ 2.6 & 5$\times$100 \\
9 & 1.5/$-5.5$/$-8$/$-11$/$-13$ & 1.23 $\rightarrow$ 1.39 & 27.4 $\rightarrow$ 2.7 & 5$\times$100 \\
10 & 4/$-5$/$-8$/$-11$/$-13$ & 1.26 $\rightarrow$ 1.39 & 26.9 $\rightarrow$ 2.3 & 5$\times$100 \\
11 & 3/$-6$/$-11$/$-12$/$-15$ & 1.41 $\rightarrow$ 1.57 & 27.3 $\rightarrow$ 2.7 & 5$\times$100 \\
12 & 3/$-2$/$-10$/$-13$/$-16$ & 1.45 $\rightarrow$ 1.61 & 26.6 $\rightarrow$ 2.8 & 5$\times$100 \\
\bottomrule
\end{tabular}
\end{table*}

\subsection{Fuel power, mixture strength and passive enleanment}
\label{sec:lambda}

The commanded fuel power was
\begin{equation}
 \dot Q_{f}=\dot m_{\mathrm{H_2}}\,\mathrm{LHV}_{\mathrm{H_2}}+\dot m_{\mathrm{MeOH}}\,\mathrm{LHV}_{\mathrm{MeOH}},
 \label{eq:mixedpower}
\end{equation}
where $\dot Q_f$ is the fuel power (W), $\dot m_{\mathrm{H_2}}$ and $\dot m_{\mathrm{MeOH}}$ are the target mass flow rates of hydrogen and methanol (kg s$^{-1}$), and $\mathrm{LHV}_{\mathrm{H_2}}=120$ MJ kg$^{-1}$ and $\mathrm{LHV}_{\mathrm{MeOH}}=19.9$ MJ kg$^{-1}$ are their lower heating values. Hydrogen and methanol densities of 0.0899 g L$^{-1}$ (0 $^\circ$C, 1 atm) and 0.792 kg L$^{-1}$ converted the volumetric targets to mass flows. Across the 25 data sets, the commanded fuel power deviated from the same-CR pure-hydrogen value by at most 0.0111\%; the metered hydrogen flow differed from its target by $-3.4$ to $+3.8$\%. The methanol energy fraction is $x=\dot m_{\mathrm{MeOH}}\mathrm{LHV}_{\mathrm{MeOH}}/\dot Q_f$.

The mixture-wide excess-air ratio was
\begin{equation}
 \lambda_{\mathrm{mix}}=\frac{\dot m_{\mathrm{air}}}{s_{\mathrm{H_2}}\dot m_{\mathrm{H_2}}+s_{\mathrm{MeOH}}\dot m_{\mathrm{MeOH}}},
 \label{eq:mixedlambda}
\end{equation}
where $\dot m_{\mathrm{air}}$ is the measured air mass flow, $\dot m_{\mathrm{H_2}}$ is here the measured hydrogen mass flow, $\dot m_{\mathrm{MeOH}}$ is the target methanol mass flow, and $s_{\mathrm{H_2}}=34.3$ and $s_{\mathrm{MeOH}}=6.47$ are the stoichiometric air--fuel mass ratios of hydrogen and methanol. At fixed air flow and fixed fuel power, composition alone would change $\lambda_{\mathrm{mix}}$ to
\begin{equation}
 \lambda_{\mathrm{th}}(x)=\frac{\lambda_{0}}{1-x+r_A x},\qquad r_A=\frac{s_{\mathrm{MeOH}}/\mathrm{LHV}_{\mathrm{MeOH}}}{s_{\mathrm{H_2}}/\mathrm{LHV}_{\mathrm{H_2}}}=1.1375,
 \label{eq:lambdashift}
\end{equation}
where $\lambda_{\mathrm{th}}(x)$ is this theoretical, composition-only value of $\lambda_{\mathrm{mix}}$ at methanol energy fraction $x$, $\lambda_{0}$ is the measured $\lambda_{\mathrm{mix}}$ of the pure-hydrogen state and $r_A$ is the ratio of the stoichiometric air demand per unit fuel energy of methanol to that of hydrogen. Equation~\eqref{eq:lambdashift} gives a 6.4\% decrease of $\lambda$ at $x=0.5$.

Because the air setting was not changed, the inducted air was predicted from the pure-hydrogen state assuming constant intake pressure and constant volumetric efficiency referenced to the measured intake temperature:
\begin{equation}
 \dot V_{\mathrm{air}}^{\mathrm{pred}}(x)=\left(\dot V_{\mathrm{air},0}+\dot V_{\mathrm{H_2},0}\right)\frac{T_{\mathrm{in},0}}{T_{\mathrm{in}}(x)}-\dot V_{\mathrm{H_2}}(x)-\dot V_{\mathrm{MeOH,v}}(x),
 \label{eq:airmodel}
\end{equation}
where $\dot V_{\mathrm{air}}^{\mathrm{pred}}(x)$ is the predicted air flow at methanol energy fraction $x$, $\dot V$ denotes a flow at standard conditions (0 $^\circ$C, 1 atm; L min$^{-1}$), subscript 0 denotes the pure-hydrogen state of the same CR, $T_{\mathrm{in}}$ is the measured intake temperature (K), $\dot V_{\mathrm{H_2}}$ is the measured hydrogen flow rate, and $\dot V_{\mathrm{MeOH,v}}=\dot m_{\mathrm{MeOH}}V_m/M_{\mathrm{MeOH}}$ is the methanol vapour flow for complete evaporation before inlet-valve closure, with molar volume $V_m=22.414$ L mol$^{-1}$ and molar mass $M_{\mathrm{MeOH}}=32.04$ g mol$^{-1}$. The first term implies that, at fixed intake pressure and volumetric efficiency, the total moles inducted per unit time scale with $1/T_{\mathrm{in}}$; the last two terms are the hydrogen and methanol vapour that displace air.

\subsection{Pressure processing and performance}
\label{sec:processing}

The absolute level of every cylinder-pressure trace was fixed by equating its value at inlet-valve closure ($-146\aTDC$) with the intake pressure recorded at that crank angle (pegging). Cylinder volume came from the crank--slider kinematics, using the effective compression ratio that the acquisition system saved for each data set. Integrating pressure over volume around the complete cycle gives the net indicated work, and dividing it by the swept volume gives the indicated mean effective pressure:
\begin{equation}
 W_i=\oint p\,\mathrm{d}V,\qquad \mathrm{IMEP}=\frac{W_i}{V_d},
 \label{eq:imep}
\end{equation}
in which $W_i$ (J) is the net indicated work of one cycle over the full 720$\CA$, $p$ (Pa) is the pegged cylinder pressure, $V$ (m$^3$) is the instantaneous cylinder volume and $V_d$ is the swept volume. The full-cycle indicated thermal efficiency relates the indicated power to the commanded fuel power:
\begin{equation}
 \mathrm{ITE}=100\,\frac{\overline{W_i}\,N/120}{\dot Q_f},
 \label{eq:ite}
\end{equation}
in which $\overline{W_i}$ is the net indicated work averaged over the cycles of a data set, $N/120$ converts the engine speed $N=600$ rpm into cycles per second for a four-stroke engine, and $\dot Q_f$ is the commanded fuel power of Eq.~\eqref{eq:mixedpower}. Because $\dot Q_f$ was constant within each CR group, the relative change of ITE equals that of IMEP. The coefficient of variation of IMEP (CoV) is the ratio of its standard deviation to its mean. The pumping mean effective pressure was 0.074--0.094 bar in all data sets, consistent with near-atmospheric intake and exhaust pressures, and changed by at most 0.014 bar within a CR group, so pumping work cannot account for the IMEP changes reported later in Section~\ref{sec:suppression}.

\subsection{Pressure oscillation, tail and acoustic metrics}
\label{sec:knockmetrics}

For MAPO, the pegged pressure was passed through a Butterworth band-pass filter of fourth order between 4 kHz and just below the 18~kHz Nyquist frequency, and the largest absolute filtered value after ignition was taken. MPRR is the steepest pressure rise per crank-angle degree between ignition and $30\aTDC$, taken as the largest absolute first difference of the pegged pressure. Across the richest conditions of the five fixed-CA50 dry sweeps, the median of the condition-mean MAPO values was 0.493 bar; rounded to 0.5 bar, this value serves as an operational reference for the condition-mean MAPO of the high-severity boundary region. The upper tail was described by the 95th percentile (P95), by CVaR95, the average of the highest 5\% of cycles, and by the share of cycles exceeding both the MAPO and the MPRR P95 of the same-CR 0\% MeOH data set.

For each cycle, a Welch spectrum of the pegged pressure was computed from ignition to $40\aTDC$ (256-sample Hann window, 75\% overlap, 2048-point transform). The dominant frequency was the frequency of maximum power spectral density within 4--15 kHz, and the oscillation energy was the spectral density integrated over the same band, i.e.\ the mean-square oscillation pressure in that band (bar$^2$). Resonance peaks were identified in the cycle-median spectrum of each data set within 5--8, 9--12 and 12--15 kHz. For a cylindrical chamber of bore $B$ \cite{Draper1938pressure},
\begin{equation}
 f_{mn}=\frac{\alpha_{mn}\,c}{\pi B},
 \label{eq:mode}
\end{equation}
where $f_{mn}$ is the frequency of the mode with $m$ nodal diameters and $n$ nodal circles, $c$ is the speed of sound of the oscillating gas, $B=82.55$ mm, and $\alpha_{mn}$ is the corresponding root of the derivative of the Bessel function: $\alpha_{10}=1.841$, $\alpha_{20}=3.054$, $\alpha_{01}=3.832$ and $\alpha_{30}=4.201$. The frequency ratios $f_{20}/f_{10}=1.659$, $f_{01}/f_{10}=2.081$ and $f_{30}/f_{10}=2.282$ depend only on geometry and identify the modes independently of $c$. A speed of sound of 800--1100 m s$^{-1}$ places $f_{10}$ at 5.68--7.81 kHz. The speed of sound of an ideal gas is
\begin{equation}
 c=\left(\frac{\gamma R_u T}{M}\right)^{1/2},
 \label{eq:soundspeed}
\end{equation}
where $\gamma$ is the ratio of specific heats, $R_u$ the universal gas constant, $T$ the gas temperature and $M$ the mean molar mass. It was computed at frozen composition for the unburned gas at the CA50 state of each data set (Section~\ref{sec:kinmethod}) and for the burned gas formed from it by adiabatic, constant-pressure equilibrium combustion; the calculation uses thermodynamic data only. To separate the causes of the change from the pure-hydrogen state to methanol energy fraction $x$, the burned-gas value of the pure-hydrogen state was recomputed at its CA50 pressure with only the intake temperature, $\lambda_{\mathrm{mix}}$ or the fuel composition set to its value at $x$, as for the flame speed in Section~\ref{sec:kinmethod}.

\subsection{Combustion progress and oscillation timing}
\label{sec:timing}

The crank angles at which 10, 50 and 90\% of the charge had burned (CA10, CA50, CA90) were obtained for every cycle from the cumulative single-zone apparent heat release,
\begin{equation}
 \frac{\mathrm{d}Q}{\mathrm{d}\theta}=\frac{\gamma}{\gamma-1}\,p\,\frac{\mathrm{d}V}{\mathrm{d}\theta}+\frac{1}{\gamma-1}\,V\,\frac{\mathrm{d}p}{\mathrm{d}\theta},
 \label{eq:hrr}
\end{equation}
where $Q$ is the apparent heat release (J), $\theta$ is the crank angle and $\gamma$ is a temperature-dependent ratio of specific heats for hydrogen--air mixtures. The model does not include the methanol contribution to $\gamma$, so CA10, CA50 and CA90 serve only as timing markers within each cycle. Oscillation onset was the first crank angle after ignition at which the 4--15 kHz Hilbert envelope of the pressure exceeded the pre-spark median by eight scaled median absolute deviations for five consecutive samples. The oscillation onset, the crank angle of MPRR and the crank angle of MAPO were then referenced to the same cycle's CA10, CA50 and CA90.

\subsection{Spark margin and knock at matched phasing}
\label{sec:marginmethod}

Spark timing was re-optimized at every substitution step, so each methanol data set was then taken at an earlier spark timing than the pure-hydrogen data set of the same CR. Because knock intensity rises with spark advance for a fixed fuel composition, a methanol data set in which every knock metric remained below its pure-hydrogen value, $K(x)<K_0$, sets a lower bound on the knock-equivalent spark advance, i.e.\ on the spark advance that would bring that metric back up to its pure-hydrogen level:
\begin{equation}
 \Delta\theta_{\mathrm{KE}}(x)>\Delta\theta_{s}(x)=\theta_{s,0}-\theta_{s}(x),
 \label{eq:kesa}
\end{equation}
where $\Delta\theta_{\mathrm{KE}}$ is the knock-equivalent spark advance ($\CA$), $\Delta\theta_s$ is the spark advance applied, $\theta_{s,0}$ and $\theta_s(x)$ are the spark timings ($\aTDC$) of the pure-hydrogen state and of methanol energy fraction $x$, and $K$ denotes any of five knock metrics, namely mean MPRR and the mean, median, 95th percentile and five-highest-cycle average of MAPO, with $K_0$ its pure-hydrogen value at the same CR. One spark timing per data set gives a bound rather than a point estimate, which would require spark sweeps at fixed $x$. Because methanol also lengthened the combustion duration, part of $\Delta\theta_s$ only compensated for the slower combustion; the accompanying shift of mean CA50 gives the phasing actually gained.

Within a data set, knock also varies with combustion phasing, so the reduction was divided into a part that appears at an unchanged CA50 and a part carried by the change of mean CA50. For each CR and methanol fraction, one regression was fitted to all cycles of the pure-hydrogen and the methanol data sets together:
\begin{equation}
 \ln K_j=a+b\,\mathrm{CA50}_j+c\,D_j,
 \label{eq:matched}
\end{equation}
in which $K_j$ is the knock metric (MAPO or MPRR) of cycle $j$, $\mathrm{CA50}_j$ is its CA50 crank angle, the indicator $D_j$ equals 1 for a cycle of the methanol data set and 0 otherwise, $a$ is an intercept, $b$ is the slope of $\ln K$ versus CA50 shared by both data sets, and $c$ is the offset of the methanol data set at equal CA50. The fractional change at matched CA50 is therefore $\exp(c)-1$. With an intercept and the indicator in the model, least-squares residuals sum to zero within each data set, so the difference between the data sets in mean $\ln K$, the logarithm of the ratio of their geometric means, splits exactly into $c$ and $b\,\Delta\overline{\mathrm{CA50}}$; the second part is the contribution of the shift $\Delta\overline{\mathrm{CA50}}$ of mean CA50 from the pure-hydrogen to the methanol data set. The 95\% confidence intervals came from 2000 resamples in which ten-cycle blocks were drawn within each data set, which keeps the correlation between successive cycles. As sensitivity checks, the change at matched CA50 was recomputed with a separate slope for each data set and with only the cycles inside the CA50 range shared by both. When the two CA50 ranges did not overlap, the whole-record reduction is reported instead, as a lower bound of the reduction at matched CA50 (see Section~\ref{sec:margin} later).

\subsection{Laminar flame speed and end-gas autoignition}
\label{sec:kinmethod}

Laminar flame speeds and end-gas ignition were computed with Cantera 3.2.0 \cite{Goodwin2025cantera} for the unburned mixture of every data set. The mixture contained hydrogen, methanol vapour and air (O$_2$ + 3.76 N$_2$) at the measured $\lambda_{\mathrm{mix}}$ (Eq.~\eqref{eq:mixedlambda}) and methanol energy fraction; methanol was assumed fully evaporated before inlet-valve closure and residual gas was neglected. The unburned-gas temperature followed from isentropic, non-reacting compression along the mean measured pressure of the analysed cycles,
\begin{equation}
 s(T_u,p)=s(T_{\mathrm{IVC}},p_{\mathrm{IVC}}),
 \label{eq:isentropic}
\end{equation}
where $s$ is the specific entropy of the unburned mixture, $T_u$ is the unburned-gas temperature at the measured cylinder pressure $p$, and $T_{\mathrm{IVC}}$ and $p_{\mathrm{IVC}}$ are the temperature and pressure at inlet-valve closure. $T_{\mathrm{IVC}}$ was taken as the measured intake temperature plus 25 K to allow for heat absorption from the walls and residual gas, with 0 and 50 K as bounds. The laminar flame speed $S_L$ was that of a freely propagating, one-dimensional, planar and unstretched premixed flame with mixture-averaged transport, evaluated at the unburned-gas state of each data set's mean CA50. To separate the contributions to the change from 0 to 50\% methanol, $S_L$ of the pure-hydrogen state was recomputed at its CA50 pressure with only $\lambda_{\mathrm{mix}}$, the intake temperature or the fuel composition set to its 50\% methanol value. End-gas autoignition was assessed with the Livengood--Wu integral \cite{Livengood1955correlation},
\begin{equation}
 I=\int_{t_{\mathrm{IVC}}}^{t_{\mathrm{CA90}}}\frac{\mathrm{d}t}{\tau(T_u,p)},
 \label{eq:lw}
\end{equation}
where $t$ is time, $t_{\mathrm{IVC}}$ and $t_{\mathrm{CA90}}$ are the times of inlet-valve closure and of the mean CA90, and $\tau$ is the ignition delay of the unburned mixture in a constant-volume adiabatic reactor, defined as the time to a 400 K temperature rise and interpolated from a table over 600--1100 K and 10--60 bar; $I=1$ marks autoignition before the end of combustion. Two mechanisms developed and validated for both fuels were used and their mean is reported: the San Diego mechanism (version 2016-12-14; 59 species, 271 reactions) \cite{SanDiego2016mechanism} and the methanol mechanism of Burke et al. (173 species, 1011 reactions), which was validated against methanol ignition-delay and jet-stirred-reactor measurements \cite{Burke2016detailed}. Both mechanisms were validated against measured flame speeds of methanol and hydrogen--methanol mixtures \cite{Liang2017laminar,Xiao2020effects,Zheng2024measurements} (Supplementary Section S10). GRI-Mech 3.0 \cite{Smith1999grimech} was also evaluated but not used, because it was optimized for natural-gas combustion and not for methanol, nor for hydrogen ignition at the low temperatures and high pressures of the present unburned gas (see also Supplementary Section S8).

The response of the flame to stretch was computed at the unburned-gas state of each data set's mean CA10, when the flame kernel is small and most strongly stretched. Following the computational approach of Davis et al.\ \cite{Davis2002markstein}, symmetric twin premixed flames in an opposed-jet configuration were solved with the same mechanisms and transport model. The inlet velocity was $3S_L$, and the nozzle separation was chosen to impose Karlovitz numbers $\mathrm{Ka}=K\delta/S_L$ of 0.06, 0.10 and 0.15, where $K$ is the strain rate, taken as the largest axial velocity gradient ahead of the flame, and $\delta=(T_{\mathrm{ad}}-T_u)/\max(\mathrm{d}T/\mathrm{d}z)$ is the thermal thickness of the planar flame, with $T_{\mathrm{ad}}$ its adiabatic flame temperature and $z$ the distance normal to the flame. At such weak stretch the flame speed varies linearly with the strain rate \cite{Law2000structure}:
\begin{equation}
 S_{u,\mathrm{ref}}=S_L-L_u K,\qquad \mathrm{Ma}=\frac{L_u}{\delta},
 \label{eq:markstein}
\end{equation}
where $S_{u,\mathrm{ref}}$ is the reference flame speed, taken as the minimum axial velocity ahead of the flame, $L_u$ is the Markstein length relative to the unburned gas, obtained from a linear fit over the three strain rates, and $\mathrm{Ma}$ is the Markstein number; a negative $\mathrm{Ma}$ means that stretch accelerates the flame. A spherical kernel of radius $r$ that grows at $\mathrm{d}r/\mathrm{d}t\approx\sigma S_L$ is stretched at $K=(2/r)\,\mathrm{d}r/\mathrm{d}t$ \cite{Bradley1996burning}, so Eq.~\eqref{eq:markstein} gives the relative change of its flame speed as
\begin{equation}
 \frac{\Delta S}{S_L}=-\frac{2\sigma L_u}{r},
 \label{eq:kernel}
\end{equation}
where $\Delta S$ is the stretch-induced change in flame speed and $\sigma=\rho_u/\rho_b$ is the ratio of unburned- to burned-gas density. Because $\mathrm{Ka}=2\sigma\delta/r$ for such a kernel, the imposed strain rates correspond to kernel radii of 0.4--2.4~mm. Equation~\eqref{eq:kernel} transfers the counterflow Markstein length to a curved, expanding kernel and is used to estimate the magnitude of the stretch effect.

\subsection{Exhaust species and statistics}
\label{sec:emissions_method}

The exhaust analyser recorded 50 samples per data set. Its channels include NO$_x$, CO, CO$_2$, H$_2$O, HCHO and several hydrocarbons but no CH$_3$OH channel, and it reports volume fractions in ppm on a wet basis: the measured H$_2$O of the pure-hydrogen and 50\% methanol data sets differed by 1--8\% from the value computed for complete combustion from the air and fuel flows of Section~\ref{sec:lambda}, and the measured CO$_2$ of the 50\% methanol data sets by 2--7\% (Supplementary Section~S5). NO$_x$, HCHO and CO are reported as changes relative to the pure-hydrogen data set of the same CR group, together with their wet-basis values.

Each data set is summarized by statistics over its cycles (means, percentiles and exceedance fractions), and a trend is regarded as general when it appears at all five compression ratios. The Supplementary Material reports a dry-hydrogen response model that tests whether the dry-H$_2$ $\lambda$--spark trend can reproduce the observed MAPO reduction (Section S1), a principal-component analysis of pressure shape (Section S2), joint tail risk and event detectability (Section S3), apparent ring-down fits (Section S4), the exhaust-analyser channels (Section S5), the spark-margin and matched-phasing results at every methanol fraction (Section S6), the pre-ignition and late-autoignition cycles at CR12 (Section S7), and the mechanism comparison, the stretched-flame calculations and the results of the kinetic calculations for each data set (Section S8).

\section{Results}
\label{sec:results}

\subsection{Knock severity of the pure-hydrogen starting states}
\label{sec:dry}

\begin{figure*}[tp]
\centering
\includegraphics[width=\textwidth]{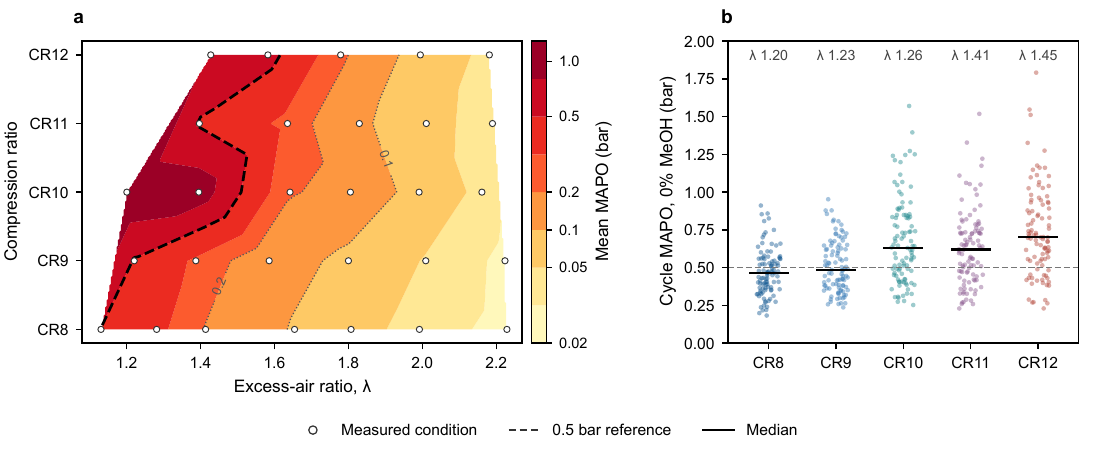}
\caption{Knock context of the substitution campaign. (a) Mean MAPO of the dry-hydrogen fixed-CA50 $\lambda$ sweeps at CR8--CR12; contours are linearly interpolated between the measured conditions (white dots), and the dashed line is the 0.5-bar operational reference. (b) Cycle MAPO of the five pure-hydrogen starting states of the substitution campaign, with medians and measured $\lambda$; the dashed line is the 0.5-bar reference; one CR12 cycle, at 3.03 bar, lies above the plotted range.}
\label{fig:context}
\end{figure*}

In the dry fixed-CA50 sweeps, mean MAPO rose steeply as $\lambda$ decreased toward the accessible rich edge of lean hydrogen operation, and only the two richest conditions at CR10 and CR12 exceeded the 0.5-bar reference (Fig.~\ref{fig:context}a). The five pure-hydrogen starting states of the substitution campaign had mean MAPO values of 0.469--0.760 bar and cycle P95 values of 0.75--1.28 bar (Fig.~\ref{fig:context}b); four lay above the 0.5-bar reference and the CR8 state was just below it. Because the knock boundary moves to leaner mixtures as CR increases (Fig.~\ref{fig:context}a), the near-boundary pure-hydrogen states at CR11 and CR12 required 25.3--25.9~L~min$^{-1}$ of hydrogen instead of 28.1--29.5~L~min$^{-1}$ at CR8--CR10; air filled the volume freed by the hydrogen, so $\lambda_{\mathrm{mix}}$ started at 1.41--1.45 instead of 1.20--1.26 (Table~\ref{tab:matrix}). The starting states were thus matched in knock severity rather than in $\lambda$, and each substitution step is compared with the starting state of its own CR.

\subsection{Timing and origin of the pure-hydrogen oscillation}
\label{sec:excitation}

\begin{figure*}[tp]
\centering
\includegraphics[width=\textwidth]{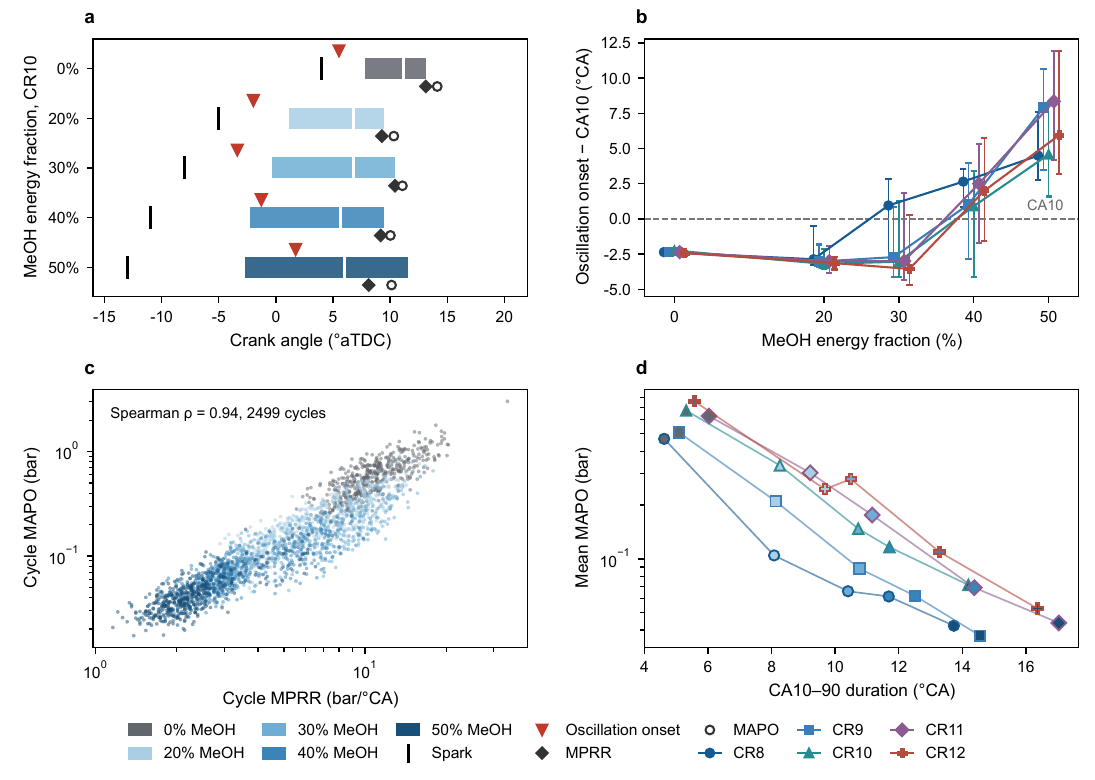}
\caption{Oscillation excitation and burn rate. (a) CR10 event timeline: spark timing (vertical bar), median CA10--CA90 interval (shaded bar; white tick at CA50), median oscillation onset (triangle), median MPRR angle (diamond) and median MAPO angle (circle). (b) Oscillation onset relative to CA10 (condition median and interquartile range) for detected events at CR8--CR12. (c) Cycle MAPO versus cycle MPRR for all 2499 analysed cycles, coloured by methanol energy fraction. (d) Condition-mean MAPO versus median CA10--CA90 duration; symbol shape denotes CR and fill colour the methanol energy fraction.}
\label{fig:excitation}
\end{figure*}

The timing of the oscillation within the combustion event identifies its origin. For pure hydrogen, the 4--15 kHz envelope crossed its onset threshold 0.4--1.9$\CA$ after ignition (condition medians), before CA10 in 97--100\% of the cycles at every CR and in all cycles of the most intense MAPO decile, and before CA50 in every cycle (Fig.~\ref{fig:excitation}a,b). The MPRR occurred 1.1--2.0$\CA$ after CA50, in the fastest part of the event, and the MAPO 0.5--1.3$\CA$ after CA90, when the burned volume and pressure were largest. In the most intense decile, the MPRR occurred 1.2--2.9$\CA$ after CA50 rather than late in the event, where end-gas autoignition would place it. Only one cycle, at CR12, combined a late, steep pressure rise (MPRR 0.5$\CA$ after CA90, 2.7 times the median of its data set) with a MAPO of 3.0 bar, consistent with end-gas autoignition (Supplementary Fig.~S6, blue trace); without it, the CR12 mean MAPO would be 0.737 instead of 0.760~bar, and the median would be unchanged at 0.701~bar. Across all 2499 analysed cycles, MAPO followed MPRR closely (Spearman $\rho=0.94$; 0.91--0.96 within each CR group; Fig.~\ref{fig:excitation}c). The oscillation was therefore excited from the start of combustion and grew with the rate of heat release, which identifies it as deflagration-excited rather than autoignition-driven.

\subsection{Intake temperature, air flow and mixture strength}
\label{sec:trajectory}

\begin{figure*}[tp]
\centering
\includegraphics[width=\textwidth]{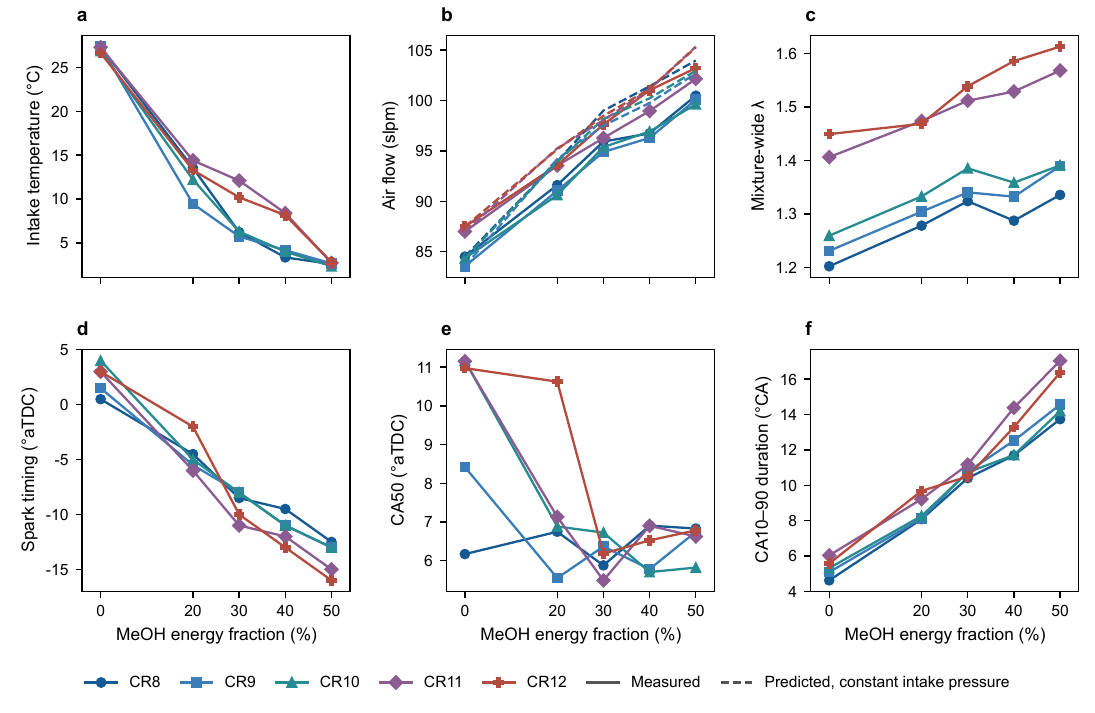}
\caption{Operating trajectory at constant commanded fuel power. (a) Intake temperature downstream of the methanol injector. (b) Measured air flow (solid) and air flow predicted by Eq.~\eqref{eq:airmodel} for constant intake pressure and volumetric efficiency (dashed). (c) Mixture-wide $\lambda$ from Eq.~\eqref{eq:mixedlambda}; the leaner starting states at CR11 and CR12 follow from the leaner knock boundary at these CRs (Section~\ref{sec:dry}). (d) Spark timing; more negative values denote earlier ignition. (e) Cycle-mean CA50. (f) Median CA10--CA90 duration.}
\label{fig:trajectory}
\end{figure*}

Methanol evaporation reduced the intake temperature from 26.6--27.4 $^\circ$C to 2.3--2.8 $^\circ$C at 50\% methanol (Fig.~\ref{fig:trajectory}a), while the controlled upstream air-heater and intake-tube temperatures and the injector-holder temperature changed by less than 3 K. With the air temperature setting and intake pressure unchanged and the air flow free to vary, the inducted air mass rose by 17.4--19.8\% (Fig.~\ref{fig:trajectory}b). Eq.~\eqref{eq:airmodel} reproduced the measured air flow at every substitution step to within 0.1--4.8\%, slightly over-predicting it, consistent with incomplete methanol evaporation before inlet-valve closure or a thermocouple reading biased low by wall-film cooling. The increase in air flow therefore follows from the colder, denser charge and from the smaller volume of hydrogen gas in the intake. Consequently $\lambda_{\mathrm{mix}}$ rose by 10.4--12.9\% (Fig.~\ref{fig:trajectory}c), although composition alone at fixed air flow would have reduced it by 6.4\% (Eq.~\eqref{eq:lambdashift}). The enleanment thus came with the methanol and was not caused separately; its effect on the combustion is separated from those of cooling and fuel replacement in Section~\ref{sec:kinetics}.

The colder, leaner charge, in which methanol had replaced part of the high-flame-speed hydrogen, burned more slowly. The CA10--CA90 duration lengthened from 4.6--6.0$\CA$ to 13.7--17.0$\CA$, 2.7--3.0 times the pure-hydrogen value (Fig.~\ref{fig:trajectory}f). Spark timing was advanced by 13--19$\CA$ (Fig.~\ref{fig:trajectory}d), which returned the late pure-hydrogen CA50 of CR9--CR12 to 5.8--6.8$\aTDC$ (Fig.~\ref{fig:trajectory}e).

\subsection{Knock metrics, IMEP and efficiency}
\label{sec:suppression}

\begin{figure*}[tp]
\centering
\includegraphics[width=\textwidth]{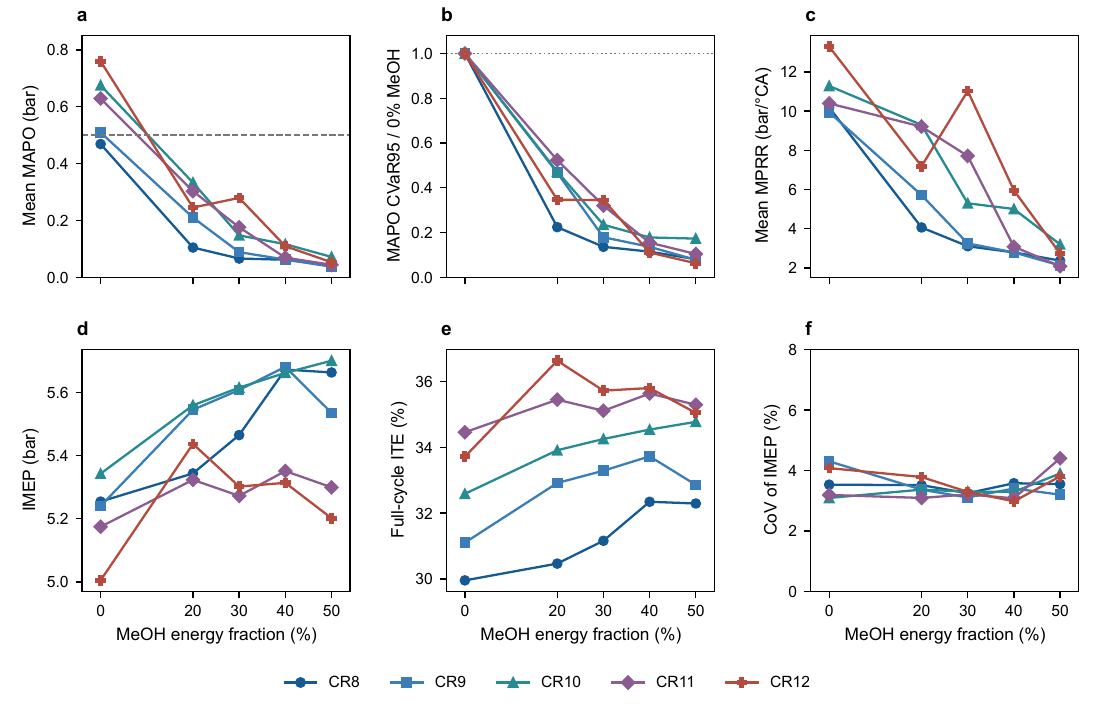}
\caption{Knock suppression and performance along the substitution path at CR8--CR12. (a) Mean MAPO; the dashed line is the 0.5-bar reference. (b) MAPO CVaR95 relative to the same-CR pure-hydrogen state. (c) Mean MPRR. (d) IMEP. (e) Full-cycle indicated thermal efficiency, Eq.~\eqref{eq:ite}. (f) Coefficient of variation of IMEP.}
\label{fig:suppression}
\end{figure*}

Mean MAPO fell by 50.7--77.7\% at 20\% methanol, 63.3--85.9\% at 30\%, 82.8--89.0\% at 40\% and 89.4--93.3\% at 50\% (Fig.~\ref{fig:suppression}a). The tail fell with the mean: MAPO CVaR95 decreased by 47.6--77.6\% at 20\% and by 82.7--93.7\% at 50\% (Fig.~\ref{fig:suppression}b); no methanol-containing cycle exceeded the same-CR pure-hydrogen MAPO P95, and no methanol-containing data set contained a cycle that jointly exceeded the pure-hydrogen MAPO and MPRR P95 (Supplementary Fig.~S3). MPRR fell by 71.6--79.9\% at 50\% (Fig.~\ref{fig:suppression}c).

\begin{figure*}[tp]
\centering
\includegraphics[width=\textwidth]{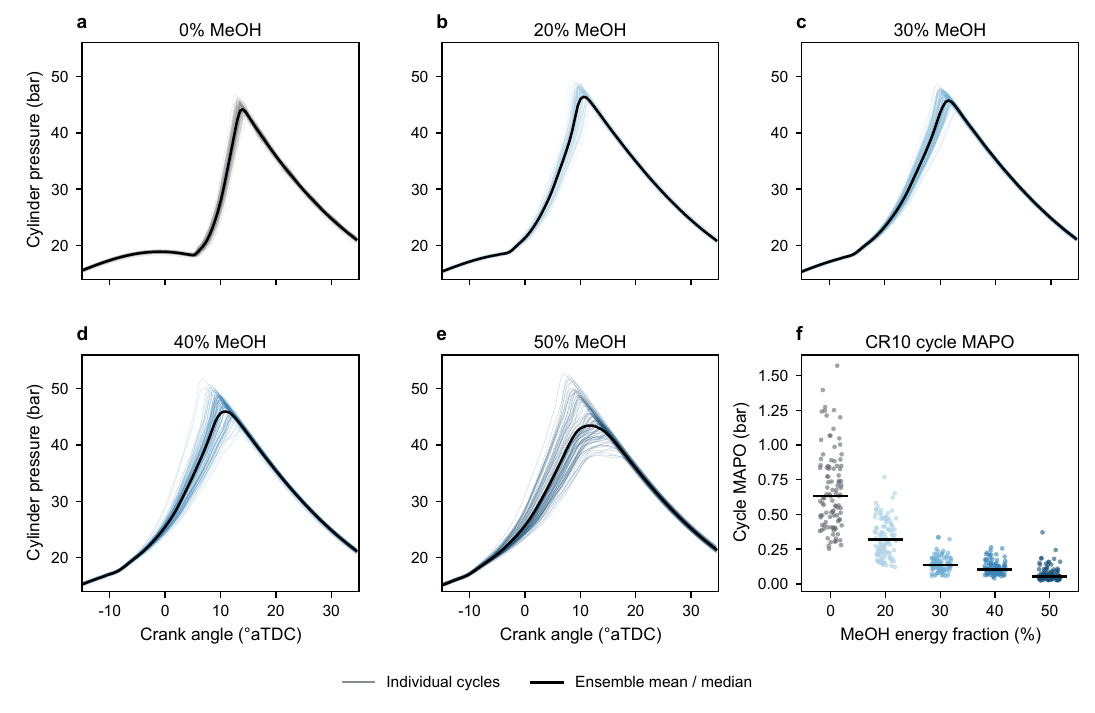}
\caption{Non-averaged CR10 cylinder pressure. (a--e) Pegged pressure of all 100 cycles (thin lines) and ensemble mean (thick line) at 0, 20, 30, 40 and 50\% methanol energy fraction, respectively. (f) Cycle MAPO with medians.}
\label{fig:cycles}
\end{figure*}

The suppression was not linked to load or efficiency. IMEP and the full-cycle ITE were higher than in the pure-hydrogen state at every substitution step and CR, by at least 1.7\% (relatively), and by 2.4--7.8\% at 50\% methanol, where ITE reached 32.3--35.3\% (Fig.~\ref{fig:suppression}d,e). The non-averaged CR10 pressure traces confirm the change in combustion character (Fig.~\ref{fig:cycles}): for pure hydrogen the pressure rose abruptly after a late start, whereas with methanol it rose gradually from before top dead centre. The slower combustion at 50\% methanol widened the spread of peak pressure between cycles, its CoV rising from 1.2--2.1\% to 4.0--9.1\%, but the CoV of IMEP remained within 3.0--4.4\% (Fig.~\ref{fig:suppression}f) and every cycle stayed in the low-MAPO cluster (Fig.~\ref{fig:cycles}f).

\subsection{Spark margin and knock at matched phasing}
\label{sec:margin}

All five knock metrics remained below their pure-hydrogen values at every substitution step and CR, although the spark timing had been advanced (Table~\ref{tab:highdose}; Supplementary Fig.~S5a,b). By Eq.~\eqref{eq:kesa}, the knock-equivalent spark advance therefore exceeded 5--9$\CA$ at 20\% methanol, 9--14$\CA$ at 30\%, 10--16$\CA$ at 40\% and 13--19$\CA$ at 50\%, and at 50\% the least-reduced metric was still only 20--28\% of its pure-hydrogen value. Part of this advance compensated for the slower combustion: at 50\% methanol, mean CA50 was 1.7--5.3$\CA$ earlier than for pure hydrogen at CR9--CR12 and 0.7$\CA$ later at CR8 (Supplementary Fig.~S5c). The engine therefore ran at a CA50 of 5.8--6.8$\aTDC$, close to the $6\aTDC$ target of the dry sweeps, with mean MAPO 89--93\% below its pure-hydrogen value; knock no longer constrained the phasing.

Comparing cycles at matched CA50 (Eq.~\eqref{eq:matched}) shows that the reduction did not follow later phasing (Supplementary Fig.~S5d--f). Inside each data set, every degree of later CA50 reduced MAPO by 8--29\% at 50\% methanol. At CR8 and CR9, where the pure-hydrogen and 50\% methanol cycles shared a CA50 range, MAPO at matched CA50 was 90.3\% (95\% confidence interval 89.6--91.2\%) and 94.0\% (93.3--94.8\%) lower, respectively, versus geometric-mean reductions of 90.8 and 92.6\% for the whole data sets. The 0.7$\CA$ later phasing at CR8 accounted for a 5\% reduction, and the earlier phasing at CR9 would on its own have increased MAPO by 24\%. At CR10--CR12 the methanol cycles burned 4.2--5.3$\CA$ earlier, and at most three pure-hydrogen cycles fell within the methanol CA50 range, so a matched value would extrapolate the slope. Because the slope was negative at every CR, with 95\% confidence intervals excluding zero, the earlier phasing alone could only have increased knock, and the reduction at a given phasing was at least the whole-record reduction of 90.4--93.3\%. Later phasing contributed to the reduction only at CR8 at 20, 40 and 50\% methanol, where mean CA50 moved 0.6--0.7$\CA$ later and reduced MAPO by at most 7\% and MPRR by at most 8\%; in the other 17 methanol data sets CA50 moved earlier, so the shift in phasing on its own would have increased MAPO (Supplementary Table~S2).

\subsection{Heat-release rate and oscillation excitation}
\label{sec:mechanism}

The excitation changed in step with the burn rate. Oscillation onset still preceded CA10 by 2.9--3.2$\CA$ at 20\% methanol and, except at CR8, by 2.7--3.6$\CA$ at 30\%, but followed CA10 by 0.9--2.6$\CA$ at 40\% and by 4.5--8.3$\CA$ at 50\% (condition medians; Fig.~\ref{fig:excitation}a,b); at 50\% only 3--13\% of detected events began before CA10. Within every CR group, mean MAPO fell by about an order of magnitude as the combustion duration lengthened from the pure-hydrogen to the 50\% state (Fig.~\ref{fig:excitation}d), and the cycles of all methanol fractions lie along the same MAPO--MPRR band (Fig.~\ref{fig:excitation}c). The 13--19$\CA$ spark advance would increase end-gas temperature and residence time at otherwise fixed conditions, yet the oscillation weakened: what changed was the rate at which the deflagration pressurized the chamber. A principal-component analysis of the low-pass pressure trace, which needs neither the knock filter nor a heat-release calculation, shows the same change independently: the component describing a late, steep pressure rise was lower at 50\% methanol than for pure hydrogen at every CR, and at CR10--CR12, where pure hydrogen burned late, the component describing phasing moved toward earlier combustion (Supplementary Section~S2). At 50\% methanol the oscillation approached the detection limit: in 9--35\% of cycles it never rose above the pre-spark timing noise, so the timing statistics rest on the remaining 65--91\% (Supplementary Fig.~S3f).

\subsection{Laminar flame speed, stretch response and end-gas autoignition}
\label{sec:kinetics}

\begin{figure*}[tp]
\centering
\includegraphics[width=\textwidth]{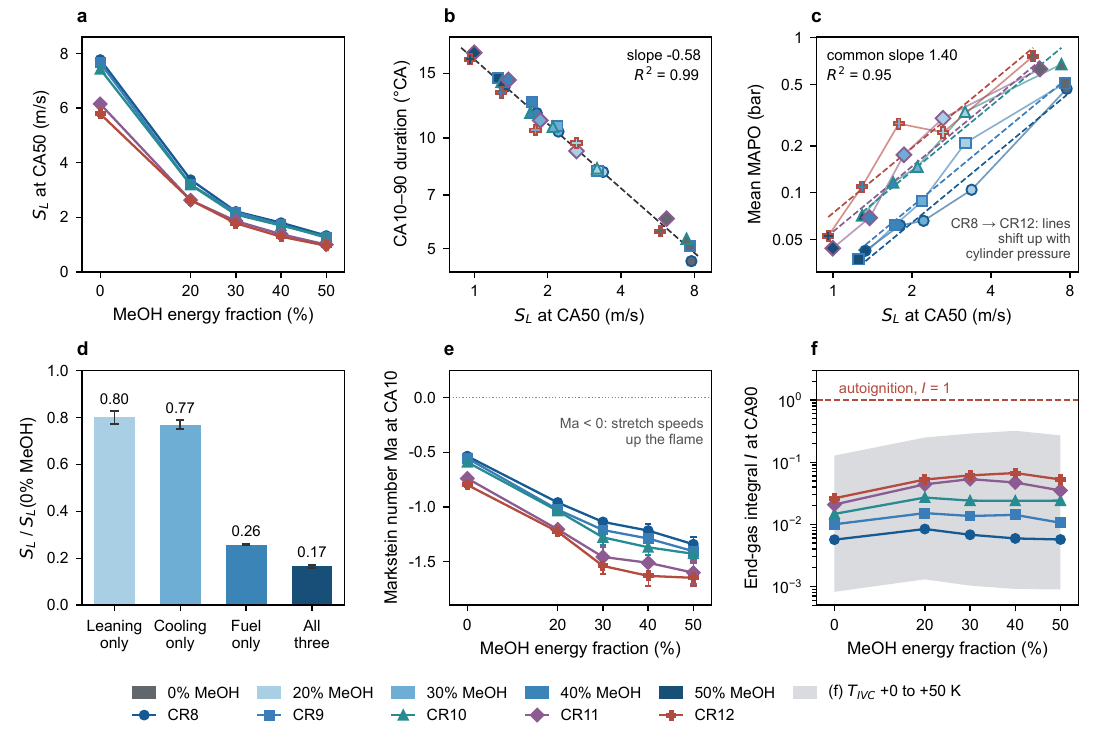}
\caption{Chemical-kinetic interpretation of the substitution at CR8--CR12. Laminar flame speeds $S_L$ and end-gas ignition delays were computed using the San Diego and Burke et al.\ methanol mechanisms for the unburned gas of each data set, compressed isentropically from inlet-valve closure along the mean measured pressure (Eq.~\eqref{eq:isentropic}); plotted values are the mean of the two mechanisms. (a) $S_L$ at the unburned-gas state of CA50; the two mechanisms differed by at most 0.18~m~s$^{-1}$, less than the symbol size. (b) Measured median CA10--CA90 duration versus $S_L$ for all 25 data sets, with a power-law fit. (c) Measured mean MAPO versus $S_L$; thin lines connect the data sets of each CR and dashed lines are a fit with a common exponent and CR-specific intercepts. (d) $S_L$ of the 50\% methanol energy fraction state relative to the pure-hydrogen state when only $\lambda_{\mathrm{mix}}$, the intake temperature or the fuel composition is changed, and when all three change; bars are the mean of CR8--CR12 and whiskers their range. (e) Markstein number at the unburned-gas state of CA10 from twin counterflow flames, Eq.~\eqref{eq:markstein}; error bars span the two mechanisms, and negative values mean that stretch accelerates the flame. (f) Livengood--Wu integral of the end gas at CA90, Eq.~\eqref{eq:lw}, for an inlet-valve-closure temperature 25 K above the intake temperature; the grey band spans the 0 and 50 K bounds over all CRs. In all panels, symbol shape denotes CR and fill colour the methanol energy fraction.}
\label{fig:kinetics}
\end{figure*}

Methanol reduced the computed laminar flame speed at the CA50 state from 7.4--7.8 m/s at CR8--CR10 and 5.8--6.2 m/s at the leaner CR11--CR12 for pure hydrogen to 0.96--1.33 m/s at 50\% methanol, 0.16--0.17 of the pure-hydrogen value (Fig.~\ref{fig:kinetics}a). The two mechanisms agreed within $\pm$8\%. The assumed inlet-valve-closure temperature shifted the absolute $S_L$ by up to $-$26\% and $+$33\% for the 0 and 50 K bounds but left the ratio between 50\% methanol and pure hydrogen at 0.16--0.18. The measured CA10--CA90 duration of all 25 data sets followed one power law of $S_L$, with an exponent of $-0.58$ and $R^2=0.99$, regardless of CR (Fig.~\ref{fig:kinetics}b). Mean MAPO fell monotonically with $S_L$ within every CR group, with a common exponent of 1.40 ($R^2=0.95$; Fig.~\ref{fig:kinetics}c). The CR-specific offsets increased from CR8 to CR12 in the order of cylinder pressure at CA50; adding that pressure to a single fit increased $R^2$ from 0.76 to 0.97, so flame speed and cylinder pressure together account for mean MAPO in all 25 data sets.

Changing one factor at a time from the pure-hydrogen state showed where the slowdown came from (Fig.~\ref{fig:kinetics}d). The 50\% methanol $\lambda_{\mathrm{mix}}$ alone reduced $S_L$ to 0.80 of the pure-hydrogen value and the 50\% methanol intake temperature alone to 0.77, whereas the fuel composition alone reduced it to 0.26; with all three changes it fell to 0.17. On a logarithmic scale, fuel replacement accounted for 74\% of the reduction, cooling for 14\% and enleanment for 12\%. At an unchanged unburned-gas temperature, fuel replacement alone still reduced $S_L$ to 0.30, so its effect arises in the chemistry and transport of the flame rather than through a lower compression temperature.

Methanol did not remove the stretch sensitivity of the lean hydrogen flame (Fig.~\ref{fig:kinetics}e). The Markstein number at the CA10 state was $-0.54$ to $-0.80$ for pure hydrogen and became more negative with methanol, reaching $-1.34$ to $-1.65$ at 50\%; the two mechanisms differed by at most 0.21. The Markstein length was $-7$ to $-8$~$\mu$m for pure hydrogen and $-25$ to $-31$~$\mu$m at 50\% methanol. At 50\% methanol, hydrogen still made up 72\% of the fuel molecules, and its fast diffusion kept the diffusion-weighted fuel Lewis number at 0.40--0.45, although methanol on its own has a Lewis number of 1.6--2.0 (Supplementary Section~S8). By Eq.~\eqref{eq:kernel}, stretch increased the flame speed of a kernel of 2~mm radius by 2\% for pure hydrogen and by 8--9\% at 50\% methanol, which is small versus the reduction of $S_L$ at the same state to 0.16--0.18 of its pure-hydrogen value. The later oscillation onset at high methanol fractions (Section~\ref{sec:mechanism}) therefore follows from the lower flame speed itself.

The end gas remained far from autoignition (Fig.~\ref{fig:kinetics}f). For pure hydrogen the Livengood--Wu integral reached only 0.006--0.026 by CA90, and at most 0.13 for an inlet-valve-closure temperature 50 K higher. Methanol did not reduce it. At equal temperature and pressure, both mechanisms gave the 50\% methanol mixture a 3--12 times shorter ignition delay than pure hydrogen at 850--950 K and 30--45 bar; the colder charge partly offset this, and the integral at CA90 was 0.006--0.067 with methanol, 1.0--2.6 times its pure-hydrogen value, while MAPO fell by up to 93\%. No data set exceeded $I=0.32$ even for the 50 K bound.

\subsection{Acoustic modes of the combustion chamber}
\label{sec:acoustics}

\begin{figure*}[tp]
\centering
\includegraphics[width=\textwidth]{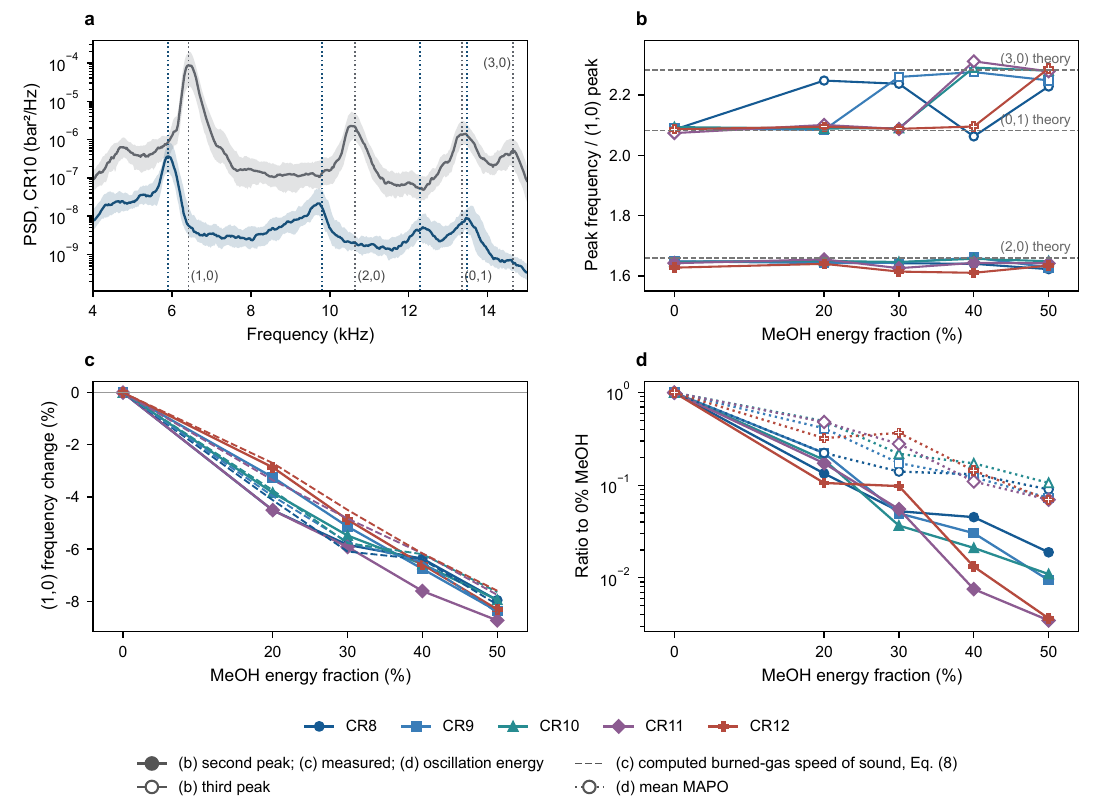}
\caption{Acoustic modes. (a) CR10 cycle-median pressure spectra at 0\% (grey) and 50\% (blue) methanol with interquartile bands; dotted lines mark the (1,0), (2,0), (0,1) and (3,0) frequencies predicted from the measured (1,0) peak by Eq.~\eqref{eq:mode}. (b) Ratio of the second (filled) and third (open) resonance peak to the (1,0) peak, with the theoretical ratios. (c) Change of the (1,0) frequency (solid) and of the computed burned-gas speed of sound, Eq.~\eqref{eq:soundspeed} (dashed). (d) Oscillation energy (solid) and mean MAPO (dotted) relative to the pure-hydrogen state. Spectral amplitudes are internally comparable but not corrected for the transfer function of the pressure-measurement chain.}
\label{fig:acoustics}
\end{figure*}

The CR10 pure-hydrogen spectrum contains four resonance peaks whose positions match Eq.~\eqref{eq:mode} (Fig.~\ref{fig:acoustics}a). The ratio of the second to the first peak frequency was 1.61--1.66 in all 25 data sets, versus 1.659 for the (2,0) and (1,0) modes (Fig.~\ref{fig:acoustics}b), so the resonator and its first two modes were the same throughout. The (1,0) frequency was 6.13--6.63 kHz for pure hydrogen and decreased by 7.9--8.7\% at 50\% methanol; the cycle-median dominant frequency decreased by 7.92--9.04\%. The speed of sound inferred from the (1,0) frequency by Eq.~\eqref{eq:mode}, 792--933~m~s$^{-1}$, lay between the computed values of the unburned gas, 563--660~m~s$^{-1}$, and of the burned gas, 925--1043~m~s$^{-1}$, closer to the burned gas. The computed burned-gas speed of sound (Eq.~\eqref{eq:soundspeed}) decreased by 7.6--8.4\% at 50\% methanol and followed the measured (1,0) frequency at every methanol fraction within 1.4 percentage points (Fig.~\ref{fig:acoustics}c), whereas that of the unburned gas decreased by 11.7--13.9\%, because methanol vapour raised its mean molar mass by 15--17\%. Of the burned-gas decrease, fuel replacement accounted for 4.5--5.0 percentage points: the CO$_2$ in the products raised the mean molar mass by 4.6--5.6\% and the flame temperature fell by 114--122 K. Enleanment accounted for 2.6--3.1 points through a 108--131 K lower flame temperature. Intake cooling accounted for 0.8--0.9 points, because the 24--25 K cooler intake lowered the burned-gas temperature by only 41--46 K (Supplementary Section S9).

The strongest peak in the 12--15 kHz band changed mode. Its ratio to the (1,0) peak was 2.07--2.09 for pure hydrogen, matching the first radial (0,1) mode, and 2.23--2.29 at 50\% methanol, matching the (3,0) mode (Fig.~\ref{fig:acoustics}b). Methanol therefore altered how the chamber was excited, and not only how strongly. The oscillation energy fell by 77.8--89.4\% at 20\% methanol and by 98.1--99.7\% at 50\%, much more than for MAPO (Fig.~\ref{fig:acoustics}d), as expected for a quantity that scales with the square of the oscillation amplitude and with its duration.

\begin{table*}[tp]
\caption{Response at 50\% methanol energy relative to the same-CR pure-hydrogen state. Spark adv.\ is the spark advance applied, a lower bound of the knock-equivalent advance (Eq.~\eqref{eq:kesa}); matched red.\ is the MAPO reduction at matched CA50 (Eq.~\eqref{eq:matched}); osc.\ energy red.\ is the reduction of the 4--15 kHz oscillation energy (Section~\ref{sec:knockmetrics}). ITE is the full-cycle value at 50\% methanol; its relative change equals that of IMEP. The $\Delta$NO$_x$ and $\Delta$HCHO columns are changes of the wet-basis volume fraction, not of emission indices. $^{a}$CA50 ranges barely overlapped; the whole-record geometric-mean reduction is given as a lower bound.}
\label{tab:highdose}
\centering
\scriptsize
\setlength{\tabcolsep}{3.2pt}
\begin{tabular}{ccccccccccccc}
\toprule
CR & \shortstack{MAPO\\red. (\%)} & \shortstack{Spark adv.\\($^\circ$CA)} & \shortstack{Matched\\red. (\%)} & \shortstack{MPRR\\red. (\%)} & \shortstack{CA10--90\\ratio} & \shortstack{Onset $-$\\CA10 ($^\circ$CA)} & \shortstack{$\Delta$IMEP\\(\%)} & \shortstack{ITE\\(\%)} & \shortstack{Freq.\\drop (\%)} & \shortstack{Osc. energy\\red. (\%)} & \shortstack{$\Delta$NO$_x$\\(\%)} & \shortstack{$\Delta$HCHO\\(\%)} \\
\midrule
8  & 90.98 & 13 & 90.3 & 76.67 & 2.97 & 4.5 & 7.78 & 32.3 & 7.96 & 98.11 & $-48.81$ & 500.40 \\
9  & 92.70 & 14.5 & 94.0 & 78.58 & 2.86 & 7.9 & 5.63 & 32.9 & 8.92 & 99.06 & $-65.69$ & 444.46 \\
10 & 89.41 & 17 & $\geq$90.4$^{a}$ & 71.62 & 2.67 & 4.6 & 6.69 & 34.8 & 7.92 & 98.90 & $-52.65$ & 435.60 \\
11 & 93.03 & 18 & $\geq$93.3$^{a}$ & 79.90 & 2.82 & 8.3 & 2.41 & 35.3 & 9.04 & 99.65 & $-84.10$ & 479.71 \\
12 & 93.07 & 19 & $\geq$93.1$^{a}$ & 79.39 & 2.94 & 5.9 & 3.93 & 35.0 & 8.33 & 99.64 & $-85.22$ & 398.65 \\
\bottomrule
\end{tabular}
\end{table*}

\subsection{Performance and exhaust emissions at 50\% methanol}
\label{sec:emissions}

Table~\ref{tab:highdose} summarizes the 50\% methanol state. Every CR combined a large reduction of MAPO, MPRR and oscillation energy with higher IMEP and ITE. Relative to the pure-hydrogen state of the same CR, NO$_x$ decreased by 48.8--85.2\% at 50\% methanol, from 5650--6470~ppm to 835--3060~ppm, whereas HCHO increased by 399--500\%, from 6.9--15.5~ppm to 37.5--77.4~ppm, and CO increased by 4--180\%, from 147--441~ppm to 285--460~ppm (wet basis; Supplementary Table~S1).

\section{Discussion}
\label{sec:discussion}

\subsection{Type of hydrogen knock near the knock boundary}
\label{sec:disc_knock}

The timing evidence places the pure-hydrogen oscillation in the class that Szwaja and Naber called light hydrogen knock, generated by fast and unstable combustion from the start of combustion with pressure-pulsation amplitudes of about 0.2--1.0~bar, as opposed to heavy knock from autoignition of the unburnt mixture above 2.0~bar \cite{Szwaja2013dual}. The mean MAPO of the starting states, 0.469--0.760~bar (47--76~kPa), corresponds to their averaged peak pressure pulsation and lies within the light-knock range. Poursadegh et al. likewise found that hydrogen near stoichiometric conditions reached standard knock intensity in a RON-rating engine without the rapid pressure rise that marks end-gas autoignition, whereas leaner mixtures autoignited but gave modified RON values of 117 at $\lambda=1.5$ and above 120 at $\lambda=2$ \cite{Poursadegh2023autoignition}. Deflagration-based knock grows from spark timing through a coupling between pressure waves and heat release in fuels with high flame speeds, such as methanol and hydrogen \cite{Singh2024deflagration}, resonance excitation scales with the local pressure gradient created by heat release \cite{Guardiola2018cylinder}, and in a hydrogen direct-injection CFR engine flame speed and combustion instability excited oscillations alongside end-gas autoignition \cite{Wang2026characterization}. The present onset timing, the MPRR at CA50, and the single MAPO--MPRR relation across 2499 cycles together show that, at these boundary states, the excitation was the deflagration. The kinetic calculations support this independently: by CA90 the end-gas Livengood--Wu integral of the pure-hydrogen states reached only 0.006--0.026, so the end gas could not autoignite while the oscillation grew (Section~\ref{sec:kinetics}). Methanol suppressed the oscillation although its chemistry did not lengthen the end-gas ignition delay, which shows that the suppression acted on the deflagration rather than on autoignition. A practical consequence of this is that hydrogen knock control near the boundary is a heat-release-rate problem: any measure that slows the pressure rise suppresses the oscillation whether or not it delays autoignition.

\subsection{Roles of fuel replacement, cooling and enleanment}
\label{sec:disc_cooling}

The computed flame speed reproduces the slower burn rate: one exponent, $-0.58$, relates the CA10--CA90 duration of all 25 data sets to $S_L$ (Fig.~\ref{fig:kinetics}b), so the three routes of the hypothesis can be weighed by their effect on $S_L$. Fuel replacement dominated. Replacing half of the hydrogen energy with methanol reduced $S_L$ to about a quarter at fixed $\lambda_{\mathrm{mix}}$ and intake temperature, whereas the measured cooling and enleanment each removed only about a fifth (Fig.~\ref{fig:kinetics}d). The cooling is nevertheless real: the 24--25 K agrees in sign and order with the evaporative cooling measured and modelled for intake methanol injection on a CFR engine, where most of the cooling came from the liquid film on the intake walls \cite{Pu2024modelling}, and is consistent with methanol's 1100~kJ~kg$^{-1}$ heat of vaporization \cite{Verhelst2019methanol}. The enleanment was not imposed: at unchanged intake pressure, a colder charge with less hydrogen gas inducted more air, as Eq.~\eqref{eq:airmodel} reproduces within 5\%. Methanol did not act by stabilizing the flame front. Lean hydrogen flames are thermodiffusively unstable and accelerate as they wrinkle \cite{Verhelst2009hydrogen}, as observed in spherical lean hydrogen--air flames up to 1.0~MPa \cite{Bradley2007laminar}. The substituted mixtures kept this behaviour, and their Markstein numbers became more negative (Fig.~\ref{fig:kinetics}e), because hydrogen still supplied most of the fuel molecules. Had the flame crossed the chamber at a speed proportional to $S_L$, the six-fold drop in $S_L$ would have lengthened the combustion duration six-fold; however, it lengthened by only 2.7--3.0-fold, in proportion to $S_L^{-0.58}$ (Fig.~\ref{fig:kinetics}b). Turbulent wrinkling at the unchanged engine speed, and the stronger response of the methanol mixtures to stretch, made up part of the loss in $S_L$. Port methanol therefore resembles ammonia, which suppresses hydrogen knock by slowing the flame \cite{Ge2023knock,Hong2024ammonia}, and adds the charge cooling effect through which water acts \cite{Mortimer2023extending} while, unlike water, carrying energy; the cooling matters because a hotter intake reduces the hydrogen knock limit \cite{Purayil2025impact}. The supporting dry-hydrogen response model (Supplementary Section S1) indicates the same: with only CR, $\lambda$ and spark timing as inputs, it predicts that the leaner mixture and earlier spark of the substitution path would by themselves reduce MAPO by at most 19\% where the data support it (CR8 and CR9), whereas the measured MAPO fell by 59--93\%. The rest of the reduction comes from what the model does not contain, i.e.\ mostly the fuel replacement and to a lesser extent the cooling, which account for 74\% and 14\% of the computed flame-speed reduction, respectively (Section~\ref{sec:kinetics}, Fig.~\ref{fig:kinetics}d).

\subsection{Knock limit and efficiency relative to other knock-control methods}
\label{sec:disc_efficiency}

IMEP and ITE rose at every substitution step, by 2.4--7.8\% at 50\% methanol, at equal commanded fuel power. Methanol also improved the knock limit: every knock metric stayed below its pure-hydrogen value after 13--19$\CA$ of spark advance, and at matched phasing MAPO was still at least 90\% lower. At CR9--CR12, CA50 advanced from 8.4--11.2$\aTDC$ to 5.8--6.8$\aTDC$, close to the CA50 of about $5\aTDC$ at which a hydrogen engine reached its best indicated efficiency \cite{Mortimer2023extending}. Both results run counter to spark retard and lean operation, which reduce hydrogen knock at the cost of phasing or power density \cite{Hong2024progressive,Purayil2024spark}. In gasoline-dominant engines, methanol--gasoline blends increased the hydrogen knock limit by up to 42.8\% through end-gas cooling \cite{Purayil2026optimizing} but reduced brake thermal efficiency and brake mean effective pressure and increased cyclic variation \cite{Purayil2024influence}, because the alcohol reduced the heating value of the liquid fuel at fixed injection quantity.

\subsection{Acoustic modes as a diagnostic}
\label{sec:disc_acoustics}

The pure-hydrogen (1,0) frequency of 6.13--6.63 kHz agrees with the 6.4 kHz (1,0) peak reported for hydrogen knock in a port-injected CFR engine of the same bore \cite{Szwaja2007comparisons}, and the Bessel ratios of Eq.~\eqref{eq:mode} identify every peak. Because $f_{20}/f_{10}$ stayed at the theoretical value, the 8--9\% frequency decrease reflects a lower speed of sound in an unchanged resonator. The computed burned-gas speed of sound reproduces this decrease. At 50\% methanol the burned gas had a 5.4--6.6\% higher mean molar mass, because its products contain CO$_2$, and was 252--288 K cooler. Fuel replacement accounted for more than half of the decrease, enleanment for about one third and intake cooling for about one tenth: the burned-gas temperature is set mainly by the heat release, so the 24--25 K cooler intake lowered it by less than 50 K. The switch of the dominant high-order peak from (0,1) to (3,0) supports the observation that light and heavy hydrogen knock excite different mode sets, with light knock propagating mainly radially \cite{Luo2016inducing}, and shows that the excitation pattern changed as well as its strength. For knock detection, a narrow frequency window centred on the pure-hydrogen (1,0) mode would need to accommodate an 8--9\% shift as methanol is added, and amplitude-only metrics should be complemented by pressure-rise rate, which selects different extreme cycles \cite{Wang2025comparative,Rockstroh2018insights}.

\subsection{Formaldehyde emissions}
\label{sec:disc_hcho}

The formaldehyde increase agrees in direction with earlier methanol engine studies, in which formaldehyde and unburned methanol increased with methanol fraction and were governed by post-flame oxidation that depends on gas temperature and residence time \cite{Wei2009formaldehyde}. During cold start of a methanol engine, formaldehyde and unburned methanol moved in opposite directions, with improved combustion increasing formaldehyde \cite{Li2009emissions}, and in a direct-injection methanol engine added hydrogen increased the in-cylinder formation of formaldehyde as an intermediate \cite{Gong2024computational}. Its magnitude here should be confirmed with a calibrated method: FTIR formaldehyde readings from a methanol-fuelled engine exceeded GC and HPLC measurements because of spectral interference and calibration differences \cite{Geng2015comparative}, and accurate FTIR measurement of methanol exhaust requires methanol-specific calibration spectra \cite{Kraus2022adaptation,Gainey2025towards}. Calibrated formaldehyde and unburned-methanol emissions, together with aftertreatment effectiveness, must be included in any optimization of the substitution fraction.

\subsection{Limitations}
\label{sec:limitations}

Three limitations bound the conclusions. First, each operating point was recorded only once, as 100 consecutive cycles; the cycle statistics therefore describe the cycle-to-cycle variation within a data set, not the repeatability of the operating point when the same test conditions are repeated. Second, each methanol fraction was tested at a single spark timing, so the spark margin of Section~\ref{sec:margin} is a lower bound; a spark sweep at fixed methanol fraction would give an actual optimized value. Finally, with one pressure transducer in the D-1 cavity, the mode shapes across the bore cannot be mapped, and the frequency response of the pressure-measurement chain affects absolute spectral amplitudes but not frequency ratios or within-system changes. The split between cooling, enleanment and fuel replacement effects in Section~\ref{sec:kinetics} is computed, with an assumed unburned-gas temperature and the laminar flame speed as a surrogate for the turbulent burn speed; varying the methanol fraction at fixed intake temperature and fixed $\lambda$ would test it experimentally.

\section{Conclusions}
\label{sec:conclusions}

Port-injected methanol was used to replace up to half of the hydrogen energy at constant commanded fuel power in a variable-compression CFR engine operated at near-boundary pure-hydrogen states. The main findings are:

\begin{enumerate}
\item Near the hydrogen knock boundary, the limiting pressure oscillation was deflagration-excited: it began 0.4--1.9$\CA$ after ignition and before CA10 in 97--100\% of cycles, and its amplitude followed the pressure-rise rate across 2499 cycles ($\rho=0.94$).
\item Methanol cooled the intake by 24--25 K and, at unchanged intake pressure, the engine inducted 17--20\% more air, enleaning the mixture by 10--13\% without any air-flow adjustment; a constant-intake-pressure model reproduced the air flow within 5\%.
\item The colder, leaner charge, with part of the hydrogen replaced by methanol, burned 2.7--3.0 times more slowly over CA10--CA90, the oscillation onset moved 4.5--8.3$\CA$ after CA10, and at 50\% methanol replacement, mean MAPO fell by 89--93\%, MAPO CVaR95 by 83--94\% and MPRR by 72--80\%.
\item The slower combustion followed the computed laminar flame speed: the CA10--CA90 duration of all 25 data sets scaled with $S_L^{-0.58}$ ($R^2=0.99$), fuel replacement accounted for about three quarters of the reduction in $S_L$ and cooling and enleanment for the rest, and the end-gas autoignition integral stayed below 0.07 at CA90 and it was not reduced by methanol.
\item When methanol was added, every knock metric stayed below its pure-hydrogen value after 13--19$\CA$ of spark advance, so methanol was worth more than that advance; at matched CA50, MAPO was 90--94\% lower where the phasing ranges overlapped and at least 90\% lower elsewhere, so the suppression was not due to later phasing.
\item There was no derating: IMEP and full-cycle ITE rose at every substitution step, by 2.4--7.8\% at 50\% methanol, and the CoV of IMEP stayed at 3.0--4.4\%.
\item The chamber resonated in the same cylindrical modes throughout; the 8--9\% frequency decrease followed the lower speed of sound of the heavier, cooler burned gas, caused mainly by fuel replacement and enleanment, and the dominant high-order mode changed from radial to circumferential.
\end{enumerate}

With knock no longer a limiting factor, formaldehyde is the main constraint on the substitution: it rose five- to sixfold, to 38--77~ppm, while NO$_x$ emissions fell by 49--85\%, so formaldehyde and unburned methanol, together with aftertreatment system considerations, decide how much hydrogen methanol can replace.

\section*{Declaration of competing interest}
The authors declare that they have no known competing financial interests or personal relationships that could have appeared to influence the work reported in this paper.

\section*{Data availability}
The data that support the findings of this study are available from the corresponding author upon reasonable request.

\section*{Acknowledgements}
The authors gratefully acknowledge that funding and support for this work were provided by King Abdullah University of Science and Technology (KAUST) and its Clean Energy Research Platform.

\section*{Funding}
This work was supported by the baseline research funding of Prof. James W.G. Turner for KAUST's Advanced Vehicle Propulsion Laboratory.

\section*{Declaration of AI use}
The authors used AI-assisted tools only for language editing and translation and take full responsibility for the content of this article.

\bibliographystyle{elsarticle-num}
\bibliography{references}

\end{document}